\documentstyle[aps,epsf]{revtex}

\def\myfigure#1#2{{\leftskip=0.10753\textwidth \rightskip\leftskip\small
\begin{figure}\baselineskip=14pt plus 2pt minus 1pt
\centerline{#1}\nobreak\smallskip\nobreak #2\end{figure}}}
\global\firstfigfalse

\begin{document}
\title{Droplet Free Energy Functional for the Morphology of Martensites
\thanks{To appear in {\it Defects in Condensed Media}, eds. K. Krishan,
C. S. Sundar and V. Kumar (SciTech Publ. Ltd., Switzerland, 1996).}}

\author{Madan Rao$^1$ and Surajit Sengupta$^2$}

\address{$^1$Institute of Mathematical Sciences, Madras 600113, India\\
$^2$Material Science Division, Indira Gandhi Centre for Atomic 
Research, Kalpakkam 603102, India}

\date{\today}

\maketitle

\begin{abstract}
Martensites are metastable phases, possessing a characteristic morphology,
usually formed during a fast quench accross a structural
transition. We attempt to understand these morphological
features using a coarsegrained free energy functional ${\cal F}[\epsilon ;
 \Phi]$ which contains, in addition to the usual strain fields $\epsilon_{ij}$
(the `` order parameter'' for the transition), the ``vacancy'' field $\phi$ 
which arises due to the geometric mismatch at a parent-product interface.
The relaxation of this mismatch is slow compared to typical front
propagation times and hence $\phi$ is essentially frozen in the
reference frame of the growing martensite front.
Minimisation of ${\cal F}$ then automatically
yeilds typical martensite morphologies. 
We demonstrate this in two
dimensions for the square to rhombus transformation and obtain 
internally twinned martensites, which grow as thin strips, for ``hard''
martensites (e.g., Fe-based alloys) or with a `single-interface', for ``soft'' martensites
(e.g., In-Tl alloys).
\end{abstract}

\pacs{PACS: 64.60.Ht, 64.60.Ak, 81.30.Kf, 81.30.-t}

\bigskip

{\section {Introduction and Phenomenology}}

\noindent
Martensites, a generic term applied to diffusionless structural
transformations producing long-lived metastable phases with characteristic
morphologies,
are observed in a wide variety of systems\cite{NISH,ROIT}, ranging from
metals, alloys and ceramics to organic crystals and 
crystalline membranes\cite{VIRUS}. Although a substantial amount
of empirical
knowledge has been acquired over the years, a systematic theoretical study of 
the morphology and kinetics of martensites\cite{KRUM,SETH} is
lacking. Analysis is made difficult by the fact that the formation
of martensite is necessarily a non-equilibrium process. A
theory of the martensitic transformation should, therefore, be
able to describe the system for {\em all} times, $t$, and strictly 
equlibrium ($t\rightarrow~\infty$) descriptions are invalid.
In this paper, we derive from microscopics a free energy
functional capable of describing a droplet of growing martensite
at intermediate times.  We obtain several, empirically 
observed, morphological features of Martensites by minimising the 
free energy functional subject to appropiate constraints. In the
rest of this section we review relevant aspects of Martensite
phenomenology. We derive the droplet free energy functional in
Section $2$. In Section $3$ we use it to investigate the
morphology of two-dimensional martensites. Lastly, in Section
$4$ we discuss our results and conclusions. 

As a typical example of a Martensitic transformation consider
the case of steel (an alloy of iron, Fe and carbon, C). Pure Fe
at atmospheric pressures undergoes an equilibrium structural
phase transition from the {\it austenite} ($\gamma$) phase, which
has an FCC structure to a BCC, {\it ferrite}
($\alpha$) phase, at a temperature $T_c = 910^{\circ}$C.
A fast quench from $T > T_c$ to 
$T \leq T_{ms} ({\em martensite\!\!-\!\!start\,temperature}) < T_c$\,, 
produces, however, a rapidly transformed
metastable phase called the {\it martensite}. On nucleation,
martensite ``plates'' grow with a constant front velocity ($\,v\sim\!10^5 
cm\,s^{-1}$ in Fe)\cite{ONSHON}\, which is comparable to typical
acoustic speeds in solids. The plates consist of an array of twinned or slipped
BCC crystals, having well defined orientational relations with
each other and with the parent matrix.
The Miller indices of the {\it habit plane} 
(the equatorial plane of the plates), are fixed relative to the parent.
 The morphology of the plates depends on the elastic moduli of the solid,
ranging from ``lens-shaped'' discs (typically $\sim\,1 \mu m$) with a 
high aspect ratio in the case
of hard Fe-based alloys to a region bounded by a single interface in the
case of the softer
In-Tl alloys.  If the elastic energy barriers are large, the plates 
stop on colliding with other plates. The resulting pattern displays
a scale invariant size distribution of the plates\cite{KS}.

Several empirical facts suggest that martensites are {\it always
metastable} with respect to the ferrite. Firstly, the detailed
morphology of the plates is a sensitive function of the quench
history. Secondly, mechanical properties of martensites are dependent
on {\it aging} and {\it tempering} --- metallurgical processes
which consist of holding steel at elevated temperatures for specific 
periods of time to obtain required properties. Lastly,
martensites are never formed when steel is cooled sufficiently slowly,
thus there is an intimate relation between the 
kinetics of cooling and the morphology --- slow cooling 
nucleates a spherical droplet of the ferrite, while rapid
quenching produces a martensite with a well-defined
morphology. Any theory of the Martensitic transformation should
therefore include variables that change slowly 
over time scales corresponding to the nucleation front
propagation. Such typical ``slow modes''of a solid\cite{NOTANKI} 
undergoing a first-order
structural transformation are the displacement field ${\bf u}({\bf r},t)$ 
(goldstone mode), the mass and momentum densities $\rho({\bf r},t)$ and
${\bf g}({\bf r},t)$ (conservation laws) and the strain tensor 
$\epsilon_{ij}({\bf r},t)$ (the symmetry breaking order parameter).
Imagine, however, a grain of the product nucleating within a parent matrix
at time $t=0$. 

\myfigure{\epsfysize2.5in\epsfbox{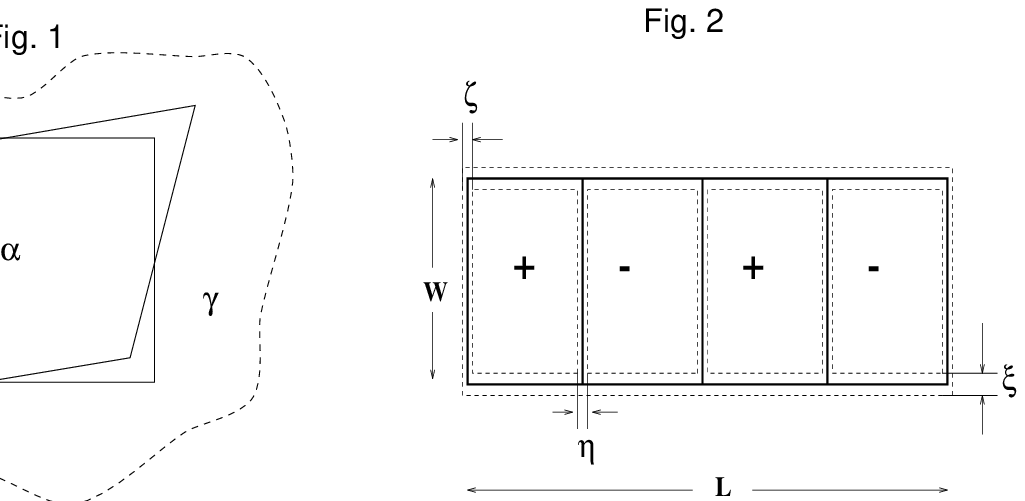}}{\vskip1cm
\noindent
Fig.\ 1 \ 
Inclusion of a product (rhombus) in a parent matrix, showing
the discontinuity in ${\bf u}$ across the parent-product interface at $t=0$.\\
Fig.\ 2 \ Variational ansatz for a polycrystal of length $L$ and width
$W$ showing $N = 4$ twins along
the $x$ axis.  The regions marked $+$ and $-$ are related to each other by
a mirror reflection at the twin interface. The interfacial  widths $\eta$, $\xi$ and $\zeta$ are also shown.}

\noindent
It is clear (Fig.\ 1), that an atomic mismatch is generated
at the parent-product interface\cite{ROIT} as soon as the nucleus is formed. 
This mismatch appears as a discontinuity in the normal component of the
displacement field across the parent-product interface\cite{LL}
(a consequence of the fact that the $\gamma \to \alpha$ structural 
transformation involves a spontaneous breaking of a  discrete symmetry),
$\Delta {\bf u} \cdot {\hat {\bf n}} \equiv \phi$, and leads to a 
local compression or dilation of the atomic environment at the interface.
Since in martensites, the transformation front velocity, $v$, is large
compared to atomic diffusion times, $\phi({\bf r},t)$ appears as a 
{\it new slow mode} (vacancy field). Once formed, this new field, relaxes 
diffusively over a time scale $\tau_{\phi} >> L/v$, where $L$ is the typical size of the grains. 
A Langevin description of the dynamics of the front, requires a 
free-energy functional {\it which describes all
intermediate configurations between an austenite and a ferrite}.
Thus the free-energy functional should include configurations corresponding
to crystalline inclusions of the $\alpha$ phase
in the $\gamma$ phase, where the vacancy field $\phi$ appears explicitly.
The usual elastic free-energy functional of a solid, $F_{el}$, has to
be augmented by an interfacial term\cite{LL}, $F_{int}$, 
describing the parent-product interface. It must be stressed that 
these droplet configurations do not affect equilibrium behaviour.
Thus the equilibrium ferrite will
be obtained by minimising $F_{el}$. The dynamics, however, is crucially 
altered by the inclusion of such fluctuations. 
In the next section we derive, starting from a
microscopic density functional theory, a free energy functional
which includes such droplet fluctuations.

\bigskip
{\section {The Droplet Free Energy Functional}}

\noindent
Consider a droplet of the product phase of size $L$ within a
parent matrix bounded by an  interface of 
thickness $\xi$. The presence of the droplet results in a strained
parent crystal such that the strain varies from zero far away
from the droplet to a limiting value at its center. The free-energy 
functional\cite{RY} of this strained parent is given by,
\begin{equation}
\frac{{\cal {F}}}{k_{B}T} = \sum_{\{{\bf R},{\bf R}^{\prime}\}} c\,
(|{\bf R}-{\bf R}^{\prime}+{\bf u}({\bf R})-{\bf u}({\bf R}^{\prime})|)\,\,,
\label{eq:FREE}
\end{equation}
where $\{{\bf R}\}$ represent the lattice vectors of the parent crystal,
${\bf u}({\bf R})$ are the displacement fields (deviation of each atom from 
the (perfect) lattice positions of the parent), and
$c(|{\bf r} - {\bf r}^{\prime}|)$ is the 
direct correlation function\cite{HM} of the liquid at freezing. 
The above expression is exact at $T=0$\,; corrections 
are of the order of the r.m.s. fluctuations of the atoms about their
perfect lattice positions, which are small in the solid phase.

Since the range of $c$ is of order $\xi$, 
the sum over $\{{\bf R},{\bf R}^{\prime}\}$ in Eq.\ \ref{eq:FREE}, can be 
split into three parts --- 
(i) ${\bf R} \wedge {\bf R}^{\prime} \,\,\,\in \,\,parent$\,,
(ii) ${\bf R} \wedge {\bf R}^{\prime} \,\,\,\in \,\,product$ and
(iii) ${\bf R} \,\,\,\in \,\,parent$ and 
${\bf R}^{\prime} \,\,\,\in \,\,product$. 
Expanding ${\bf u}({\bf R}^{\prime})$ about ${\bf R}$ in the first
two parts, leads to the usual bulk elastic free-energy functional 
$F_{el}$. However, as argued above, the assumption of continuity
breaks down in (iii), and so {\it such an expansion cannot be carried out}.
In the limit $\xi/L \ll 1$, the interface can be parametrized by a sharp 
surface 
$\Gamma\,({\bf r})=0$, with ${\bf R}$ and ${\bf R}^{\prime}$ lying
infinitesimally close to $\Gamma =0$.
Across this interface, the normal component of ${\bf u}$ is 
discontinuous\,; $lim_{\left|{\bf R}-{\bf R}^{\prime}\right|
\to 0} ({\bf u}({\bf R}) - {\bf u}({\bf R}^{\prime}))\cdot
{\hat {\bf n}} \equiv ({\bf u}_{+} - {\bf u}_{-})\cdot{\hat {\bf n}} = \phi\,
\delta(\Gamma)$, where ${\hat {\bf n}}$ is the unit normal to the
interface. The discontinuity $\phi({\bf r}) = (\rho({\bf r})|_{\Gamma=0} 
- \rho_{av})/\rho_{av}$ is the local vacancy field, 
while $\rho$ is the local density and  $\rho_{av}$ is the density averaged 
over a unit cell of the parent. In the continuum limit, the contribution to 
${\cal {F}}$ from (iii), to leading order in the discontinuity, reduces to
\begin{equation}
\frac{\gamma}{2\xi^2} \int \,d{\bf r}\,\, [\,({\bf u}_{+} - {\bf u}_{-}\,)
\cdot{\hat {\bf n}}\,]^2 \,\,\delta(\Gamma)\,\,.
\label{eq:INT}
\end{equation} 
The coefficient $\gamma \equiv \int \,d{\bf r} \, \delta(\zeta)
 \xi^2 c^{\prime \prime}(r)$ is clearly the inverse
{\it surface compressibility modulus} of vacancies, $\phi$. 
This is expected to be of
the same order of magnitude as the other elastic moduli, and thus ranges 
from ${\cal O}(10^{-3})$ for soft materials (e.g., In-Tl) to ${\cal O}(1)$
for hard materials (e.g., Fe-Ni).
Since the strains $\epsilon_{ij} \equiv (\partial_i u_j + \partial_j u_i)/2$
 are continuous across the interface, we replace the
delta function in Eq.\ \ref{eq:INT} by a regulator 
$[1-\exp(-\xi \, \partial_{n} \epsilon_{ij})^2]$, so that the final 
regulated free-energy functional
which incorporates {\it all the slow modes} in the problem is,
\begin{eqnarray}
{\cal {F}}  =   F_{el} + F_{int} 
            =   F_{el} + \frac{\gamma}{2} \int \,d{\bf r}\,\, {\phi^{2}} \,\,
({\partial_{n}}\,\, \epsilon_{ij}\,)^2  \,\,.
\label{eq:FREEPHI}
\end{eqnarray}

At the intial time ($t=0$),
the transformed region (product) is simply obtained as a geometrical
deformation of the parent, Fig.\ 1, which fixes the initial value of $\phi$.
Having created this discontinuity $\phi$ at the parent-product interface,
it will diffuse over a time $\tau_{\phi}$. 
Note the two time scales relevant to our kinetics --- the quench rate 
${\tau}^{-1} = (dT/dt)/T$ and the 
vacancy relaxation time $\tau_{\phi}$. Accordingly, two extreme dynamical
limits, suggest themselves ---  a ``slow quench'', $\tau \gg \tau_{\phi}$,
where the $\phi$ field relaxes instantaneously\,; and a ``fast quench'',
$\tau \ll \tau_{\phi}$, where the $\phi$ field is ``frozen'' in the frame
of reference of the moving nucleation front and $\phi$ relaxes
only once the growth of the front stops\cite{NOTEIMP}. Consequences of an
infinitely {\em slow} quench is well known from standard classical
nucleation theory\cite{BEKDO} since the $\phi$ field relaxes to
zero instantaneously and hence does not contribute. In the next
section we study infinitely {\em fast} quenches which
produce martensitic morphologies in the simple setting of a
two-dimensional first order structural transformation.

\bigskip
{\section {Morphology of 2D Martensites}}

\noindent
We study the first-order structural transformation from a square (austenite)
to a rhombus (ferrite) unit cell in two dimensions\cite{TWOD}.
Our results, which can be easily extended to the tetragonal
to orthorhombic transition (essentially a 2-dim square to 
rectangular\cite{KRUM}),
 are of relevance to structural transformations in alloys like 
In-Pb, In-Tl, Mn-Fe. This transformation involves a shear+volume deformation,
and so the strain order parameter $\epsilon_{ij}$ has only one nontrivial
component ${\bf e}_3=(u_{xy}+u_{yx})/2$.

The bulk elastic free energy has three minima --- one corresponding to the 
undeformed square
cell (${\bf e}_{3}=0$) and the other two corresponding to the two variants of 
the
 rhombic cell (${\bf e}_{3}=\pm {\bf e}_{0}$). This immediately leads to the 
free-energy functional $F \equiv  F_{el} + F_{int}$ (in dimensionless 
variables),
\begin{eqnarray}
F_{el} + F_{int}  
       =    \int dx\,dy\,\, [\,\,a\,{{\bf e}_3}^2 -\,{{\bf e}_3}^4+\,{{\bf
              e}_3}^6 + 
  \{(\partial_x{\bf e}_3)^2+(\partial_y{\bf e}_3)^2\} + 
\,\,{\gamma} \,(\phi\, \partial_n{\bf e}_3)^2 \,\,\, ]\,\,.
\label{eq:FREE2D}
\end{eqnarray}
The three minima of the homogeneous part of $F$ at ${\bf e}_3=0$ (austenite)
and ${\bf e}_3\equiv\pm{\bf e}_{0} = \pm [(1+\sqrt{1-3a})/3]^{1/2}$ (ferrite),
 are obtained in the 
parameter range $0 < a < 1/3$. The parameter $a$ is the degree of undercooling
and plays the role of temperature. 
 
An infinitely fast quench implies that the 
$\phi$ field is ``frozen''  
in the frame of reference travelling with the
front, ${\bf r} \to {\bf r} - {\bf v} t$ and $\phi_{t} \to -{\bf v}\cdot
{\bf {\nabla} \phi}$, where $\phi_{t}$ denotes the derivitive with respect to
time. 
Quenching the solid across the structural transition, nucleates an inclusion
which initially grows as a ferrite. Further growth as a single-domain
ferrite is discouraged by $F_{int}$. The growing nucleus, gets around this
problem by creating a twin. This costs interfacial energy which is however
small. We use a variational strain profile to include
specifically such twinned domains.
For a start, let us consider a rectangular nucleus of length $L$
(along the $x$-axis) and width $W$ (along the $y$-axis). The nucleus
is divided into $N$ twins (Fig.\ 2) with $e_{3}$ alternating between
$e_{0}$ and $-e_{0}$. The $N-1$ twin interfaces, all of thickness $\eta$, 
can be parametrized by $e_{3}\,(x)$, which for the $i$-th interface takes 
values $-e_0$ (ferrite) at  $(i-1)L/N +\eta/2< x < iL/N-\eta/2$ and $e_0$ 
(twin ferrite) at $iL/N+\eta/2 < x < (i+1)L/N-\eta/2$, with a 
linear interpolation in between. The thickness of the 
austenite-ferrite interface
along the $x$-axis ($y$-axis) is $\zeta$ ($\xi$) within which
$e_{3}\,(x)$ interpolates linearly between $0$ and $e_0$ (or $-e_0$). 
The free-energy of a nucleus with $N$ twins is 
\begin{eqnarray}
F(L,W) & = &[\,\, \Delta F LW + (2I-\Delta F)(L \xi+W \zeta)+ 
             (N-1)(I-\Delta F) W \eta 
+ 2 {e_0}^2 (\frac {L}{\xi}+\frac {W}{\zeta}) + 
             4{e_0}^2 (N-1)\frac
{W}{\eta}\,\,]  \nonumber \\
       &   &
 + \frac {\gamma {e_0}^4}{\xi} [\,\, \frac {2L^3}{3N^2}
+ L^2 \eta (\frac {1}{N^2} - \frac {1}{N}) + 
             L {\eta}^2 (\frac {1}{2} - \frac
{3}{4N}) - \frac {L\zeta}{N} (\frac {L}{N} - \frac {\zeta}{4} - \frac 
{\eta}{2})\,\,] 
   + \frac {\gamma {e_0}^4}{\zeta} [\,\, \frac {2W^3}{3}
- W^2 \xi + \frac {W {\xi}^2}{2}\,\,]
\,\,.
\label{rect}
\end{eqnarray}
The interfacial widths are taken as variational parameters.
The minimised widths $\zeta$ and $\eta$
are independent of the number of domains $N$, while the width $\xi$
scales as $L/N$. The free energy, once minimized with respect to
the widths, is a function of $L$, $W$ and $N$. As is 
aparent from Eq.\ref{rect}, ${\cal F}$ goes through a minimum as $N$ is
varied,  for every value of $L$ and $W$. This is a consequence of the
competition between the interfacial energy at the twin interface (which
disfavours twinning) and the interfacial energy at the parent~-product 
interface which favours twinning since it reduces $\phi$. Minimising the
free energy with respect to the number of twins $N$ immediately gives
us, $L/N = W^{\sigma}$ (Fig 3).  
The exponent
$\sigma \sim \frac{1}{2}$ for small $W$, for
large $W$ there are  deviations. The exponent
$\sigma$ is empirically known\cite{KRUM} to lie between $.4$ and
$.5$. A relation similar to ours has been obtained (with
$\sigma=\frac{1}{2}$) from equilibrium continuum elastic theory\cite{KRUM}.
It should be noted here, however, that the statement we make 
is stronger --- we imply that $L/N = W^{\sigma}$ {\em for all times} during
the growth of a martensite plate as a consequence of demanding
{\em local} equilibrium. Our prediction for $L/N$ can be verified from in situ
transmission electron microscopic studies of growing martensite fronts.

The free energy minimised with respect to {\em all} the ``fast''
variables (viz. the interfacial widths and $N$) is plotted in
Fig. 4. 
One observes that for $\gamma=1.0$ (Fig
4), the free energy, for all choices of $L$, shows a minimum in
$W$ for $W<L$, so that the system will always prefer to have thin
rectangular strips reminiscent of acicular martensites seen in
Fe-Ni or Fe-C systems. For $\gamma=.001$ there is no such
minimum and the martensitic region would like to grow in both
directions.

\myfigure{\epsfysize3in\epsfbox{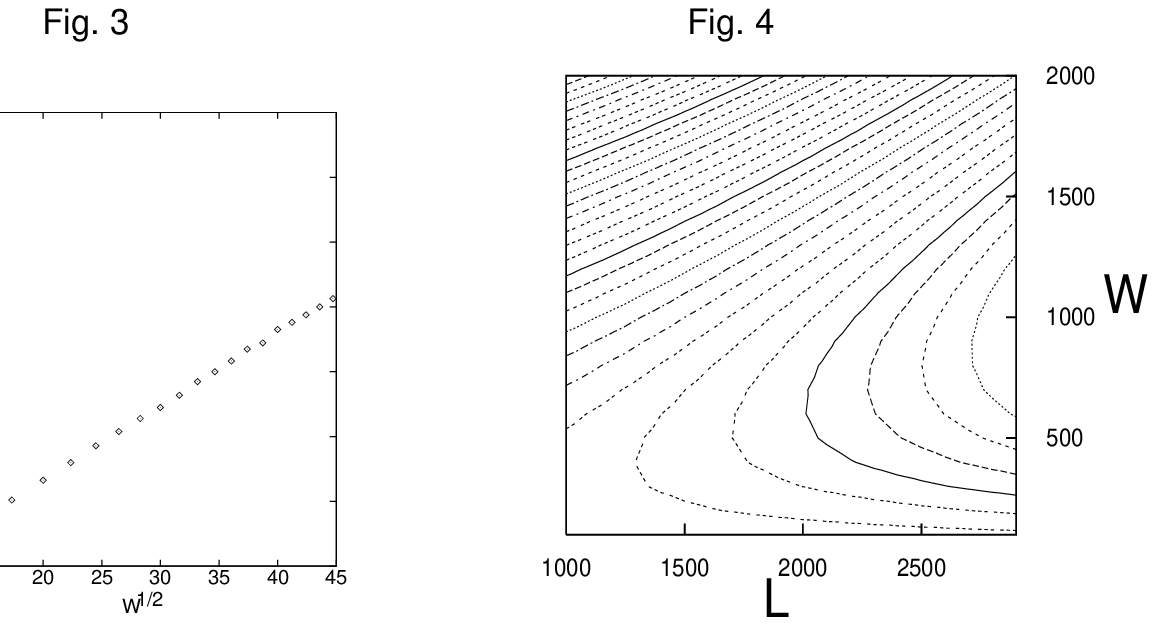}}{\vskip-0.5cm
\noindent
Fig.\ 3 \ $L/N$ plotted against $W^{1/2}$ for $L= 20000$ and $\gamma= .001$. \\
Fig.\ 4 \ Contour plots of the droplet free energy minimized with respect 
to the interfacial widths and $N$
in the $L-W$ plane for  $\gamma=1.0$. Note that there is a minimum in $W$ for all choices of $L$ (for
large $L$ and $W$) in this case.}

\noindent
Once growth in one of the directions traverses the
entire extent of the sample, growth will proceed thereafter
only along a single interface. Such single interface growth is
seen in ``soft'' In-Tl systems. The critical nucleus is obtained
by locating the saddle point in each of these free energy
surfaces. Using the quantity $\kappa = (\xi \zeta/L
W)^{1/2} $ as a measure of the sharpness of the interface (indeed one can talk
of a definite droplet shape only if the size of the droplet is larger than this measure),
we find that the critical nucleus is diffuse for both
$\gamma=1.0$ ($\kappa=.4$) and $\gamma=.001$ ($\kappa=1.43$).
However, as the size of the droplet increases $\kappa$ rapidly reduces to
zero in both cases. 

\bigskip
{\section {Summary and discussion}}

\noindent
In summary, we have derived from microscopics a free energy functional
${\cal F}[\epsilon_{ij},\phi]$
capable of describing droplets of a martensitic product enclosed
within a parent austenitic matrix. Essential to this description
is a new field $\phi({\bf r},t)$ viz. the geometric mismatch between the parent
and the product at the austenite-martensite interface. This
mismatch relaxes diffusively. For martensites, the typical front
velocities are empirically known to be very high so that
$\phi({\bf r},t)$ fails to relax within the timescale of front motion. We
have therefore minimised our droplet free energy functional, using a
variational strain profile,
subject to the constraint $\phi({\bf r},t) = \phi({\bf r},t=0)$.
We show that typical martensite morphologies drop out naturally from such
a procedure. The condition of local equilibrium yields
experimentally verifiable quantitative morphological features.
In this context, it is worth
noting that in practice, real martensites in ``hard'' systems
(eg. Fe-Ni) do not occur as rectangular strips but as lens
shaped objects. 
The following argument will show that a droplet in the form of
``convex-lens'' has, in fact, lower energy. The interfacial energy
at either end of the rectangular strip (where the interface is along the 
${\hat y}$ direction) is ${\cal O}(\,\zeta\,W(L)\,) = {\cal O}(W^2)$, where
$W(L)$ is the width of the strip at $x=L$. This can be reduced further by
making $W(L)$ (and $W(0)$, the width of the strip at $x=0$) 
as small as possible. The ``lens'' shaped 
structure thus obtained is consistent with structures seen in acicular
martensites, like Fe-Ni and Fe-C. 

\noindent
We would like to thank S.\ G.\ Mishra and  V.\ S.\ Raghunathan for useful discussions.


\begin{references}
\bibitem{NOTANKI}
The high thermal conductvity of metallic martensites, transports heat
rapidly across the transformation front and is therefore not a `slow' mode.
\bibitem{NOTEIMP}
Long range stress fields are generated when $\phi$ relaxes after the martensite
has formed (since the austenite will close any gaps generated at the
parent-product interface). No such stress fields are generated when $\phi$
relaxes much faster than the formation of a critical nucleus (and so the
ferrite formation is not accompanied by long range stress fields).
\end{references}
\end{document}